\documentclass[12pt]{article}

\usepackage{epsfig}
\usepackage{amsbsy}
\usepackage{amssymb}
\usepackage{amsmath}

\begin{document}

\title{Use of cumulants to quantify uncertainties in the HBT measurements of the homogeneity regions}
\author{K.Zalewski
\\ M.Smoluchowski Institute of Physics
\\ Jagellonian University, Cracow\footnote{Address: Reymonta 4, 30 059 Krakow,
Poland, e-mail: zalewski@th.if.uj.edu.pl. This work has been partly supported
by the Polish Ministry of Education and Science grant 1P03B 045 29(2005-2008).
}
\\ and\\ Institute of Nuclear Physics, Cracow}
\maketitle

\begin{abstract}
Let us denote by $p(\textbf{x}|\textbf{K})$ the space density of the points
where identical particles of some kind, e.g. $\pi^+$ mesons, with momentum
$\textbf{K}$ are produced. When using the HBT method to determine
$p(\textbf{x}|\textbf{K})$ one encounters ambiguities. We show that these
ambiguities do not affect the even cumulants of the distribution
$p(\textbf{x}|\textbf{K})$. In particular, the HBT radii of the homogeneity
regions, which are given by the second order cumulants, and the distribution of
distances between the pairs of production points for particles with momentum
$\textbf{K}$ can be reliably measured. The odd cumulants are ambiguous. They
are, however, correlated. In particular, when the average position $\langle
\textbf{x} \rangle(\textbf{K})$ is known as a function of $\textbf{K}$ there is
no further ambiguity.
\end{abstract}
\noindent PACS numbers 25.75.Gz, 13.65.+i \\Bose-Einstein correlations,
interaction region determination. \vspace{0.5in}

\section{Introduction}
Femtoscopy, i.e. the study of momentum correlations among particles at small
relative velocities in order to get information about the interaction regions
where the hadrons are produced, is now considered so important that workshops
devoted to correlations and femtoscopy are organized every year (see e.g.
\cite{PAP}). The HBT method is the oldest and much used tool of femtoscopy. For
reviews with hundreds of references see e.g. \cite{WIH, CSO, LIS, ZAL0}. There
is a variety of models which predict the correlation functions depending on
some parameters. Fitting these parameters to the data one obtains information
about the interaction regions. Much less attention has been devoted to the
problem: given the data what can be deduced without using a specific model. A
well-known, important result is that very little can be learned about the whole
interaction region. It is much more fruitful to concentrate on the homogeneity
regions. Homogeneity region $\textbf{K}$ is the region where hadrons with
momentum $\textbf{K}$ are produced. For a homogeneity region, much information
with little model dependence can be obtained using the imaging method (see
\cite{DAP} and references quoted there). The imaging method, however, yields
the distribution of distances between the points where identical hadrons with a
given momentum are produced. This is less information than contained in the
distribution of the production points themselves. Therefore, the question
arises: can one go any further and if not what information from models must be
added?

Let us denote $p(\textbf{x}|\textbf{K})$ the probability distribution for the
points where the hadrons (of the type considered) with momentum $\textbf{K}$
are produced in a given kind of interaction. We will call
$p(\textbf{x}|\textbf{K})$ the profile of homogeneity region $\textbf{K}$.
Barring some pathological cases, the profile can be unambiguously determined
when all its moments are known. All the moments can be unambiguously determined
when all the cumulants are known. In the present paper we concentrate on
cumulants. We find that all the even cumulants can be unambiguously determined
from the data - i.e they are measurable. This implies in particular two
important, known results. The famous HBT radii and the distribution of
distances between the production points are measurable. Actually, knowing the
distribution of distances is equivalent to knowing all the even culmulants.  A
single odd cumulant is not measurable. In particular, the $\textbf{K}$
dependence of the first order cumulants, which are equal to the components of
the the first order moments $\langle \textbf{x} \rangle(\textbf{K})$, can have
an arbitrary dependence on $\textbf{K}$. This implies, as a very special case,
the well-known result that the position of the center of the whole interaction
region is not measurable. Changing the relative positions of the homogeneity
regions, however, one usually has to change their shapes accordingly. Unless
$p(\textbf{x}|\textbf{K})$ is a Gaussian, rigid shifts of the homogeneity
regions with respect to each other are not allowed. For an example where the
skewness gets changed see \cite{BIZ}. The common knowledge is (cf. e.g.
\cite{LIS}) that the correlation functions can  provide at best the
distribution of the relative positions of particles with identical velocities.
We find that for given $\langle \textbf{x} \rangle(\textbf{K})$ the full
profiles $p(\textbf{x}|\textbf{K})$ can be measured. This result may be useful.
Macroscopic models, like the hydrodynamical ones, are more likely to predict
correctly the $\textbf{K}$-dependence of the positions of the homogeneity
regions than finer details of their shapes. According to our result, however,
function $p(\textbf{x}|\textbf{K})$ is measurable when the input includes
$\langle \textbf{x} \rangle(\textbf{K})$ for every $\textbf{K}$.

Our analysis is based on two main assumptions. One is that formula
(\ref{karczm}) is approximately valid. This can be interpreted as the
assumption that, besides the Bose-Einstein correlations, the particles are
uncorrelated. The other assumption is that formulae (\ref{emiden}) and
(\ref{emifun}) are approximately consistent.  This is known to be true when we
are sufficiently close to the classical limit. Both these assumptions hold for
a broad class of models. Our results are model independent for models within
this class.

\section{Basic formulae and assumptions}

The probability of emitting at time $t$ a particle with momentum $\textbf{K}$
from space-time point $\textbf{X}$ is given by the emission function $S(X,K)$
\cite{LIS}, \cite{PRA}. Since the particles are emitted on mass shell, we have

\begin{equation}\label{onmass}
  K^0 = E(\textbf{K}) \equiv \sqrt{m^2 + \textbf{K}^2},
\end{equation}
where $m$ is the particle mass. With this definition, the unnormalized profile
of homogeneity region $\textbf{K}$, i.e. the space distribution of points where
the identical hadrons, say $\pi^+$ mesons, with momentum $\textbf{K}$ have been
produced, is given by the integral over time

\begin{equation}\label{emiden}
  \overline{p}(\textbf{X}|\textbf{K}) = \int_{-\infty}^{+\infty}\!\!dt\;S(\textbf{X},t,K).
\end{equation}
This profile differs only by an $\textbf{X}$-independent factor from the
profile $p(\textbf{X}|\textbf{K})$, normalized to one at each $\textbf{K}$.
Therefore, either profile can be used to describe the homogeneity regions. In
this section we changed the notation for the position vector from $\textbf{x}$
to $\textbf{X}$. This has been done to improve the correspondence with the
other definition of the emission function described below. Actually, unless the
sources are point-like in space-time, $\textbf{X}$ in the following is not
quite the position vector of a particle, but our basic assumption is that the
two definitions of $S$ are, with sufficient accuracy, consistent. Then, except
in the integrands of next three formulae, $\textbf{x}$ and $\textbf{X}$ can be
used exchangeably.

On the other hand one uses \cite{LIS}, \cite{PRA} the formula
\begin{equation}\label{emifun}
  S(X,K) = \sum_F\int\!\!d^4y\;T^*_F(X+\frac{1}{2}y)T_F(X-\frac{1}{2}y)e^{iKy},
\end{equation}
where the sum is over the states of all the other particles in the system.

Following \cite{PRA} we make the approximation that each source can emit a
particle only at one time, though different sources may emit particles at
different times. This assumption of instant sources cannot be rigorously true,
it would contradict Heisenberg's uncertainty relations, but as  an
approximation it is in the spirit of the models being used. Thus e.g. Kopylov
and Podgoretsky  in their classical paper \cite{KOP} explicitly introduce
long-lived sources, but then they average over an internal, unmeasurable
parameter ($E$) which makes these sources equivalent (for the HBT analysis!) to
sets of instant sources. Many papers go even further and use sources which are,
or can be decomposed into sources, point-like in space-time \cite{LIS},
\cite{BRF}. A general derivation of the relation between correlation functions
and emission functions, which is still \cite{DAP} considered to be the most
detailed, has been given by Anchishkin, Heinz and Renk \cite{AHR}. Analyzing it
one easily finds that the proof breaks down if the sources are not, or at least
for the HBT analysis cannot be replaced by, instant sources. One also finds
\cite{ZAL4} that $S$ for coherent sources extended in time cannot be
interpreted as a phase space density of the particles produced at a given time.

For instant sources the time component $y^0$ of $y$ is zero by assumption.
Therefore, the integration over $y^0$ and the term $K_0y^0$ in the exponent
drop out. Denoting by $t_i$ the time when a source fires and using the labels
$F_i$ for the sources that fire at $t=t_i$ we get

\begin{equation}\label{}
  S(\textbf{X},t_i,K) = \int\!\!d^3y\;\sum_{F_i}T^*_{F_i}(X+\frac{1}{2}y)
  T_{F_i}(X-\frac{1}{2}y)e^{-i\textbf{K}\cdot\textbf{y}}.
\end{equation}
Usually $t_i$ is considered a continuous variable, but for comparison with the
Wigner function it is more convenient to start with discrete time variables
and, if necessary, introduce the continuous variables later.

The sum over $F_i$ yields a result proportional to the complex conjugate of the
corresponding density matrix. Denoting the (real, non-negative) proportionality
coefficients by $N(t_i)$ one gets

\begin{equation}\label{emirho}
S(\textbf{X},t_i,K) = N(t_i) \int\!\!d^3y\;\rho^*(X+\frac{1}{2}y,
X-\frac{1}{2}y) )e^{-i\textbf{K}\cdot \textbf{y}} =
N(t_i)W(\textbf{X},t_i,\textbf{K}).
\end{equation}
Actually, the integration $d^3y$ yields the complex conjugate of the Wigner
function. Since, however, the Wigner function is real the second equality is
correct. Summation over $t_i$ yields

\begin{equation}\label{emiwig}
  p(\textbf{X}|\textbf{K}) = N W(\textbf{X},\textbf{K}),
\end{equation}
where the Wigner function $W(\textbf{X},\textbf{K})$ is the average over the
Wigner functions $W(\textbf{X},t_i,\textbf{K})$ and $N$ is a
$\textbf{K}$-dependent normalization factor chosen so that for every
$\textbf{K}$ the probability distribution $p(\textbf{X}|\textbf{K})$ is
normalized to one as it should.

This result shows that the definition of the emission function as a phase space
density, used to derive formula (\ref{emiden}), is only approximately
consistent with (\ref{emifun}). A Wigner function is not a classical
distribution. It can take negative values, it is bounded by the condition
$|W(X,\textbf{K})| < \pi^{-3}$ etc. Moreover, condition (\ref{onmass}) is
automatically satisfied for the density, while for the Wigner function $K^0 =
\frac{1}{2}( p_1^0 + p_2^0)$, where both $p_1$ and $p_2$ are on mass shell, can
satisfy (\ref{onmass}) only approximately for small $|\textbf{q}|$.  Using the
Wigner function for the phase space distribution can be justified in the
classical limit \cite{HIL}. Crude estimates (cf. e.g. \cite{ZAL2}) suggest that
it is a very good approximation for the homogeneity regions in heavy ion
scattering, though not necessarily so for much smaller interaction regions as
e.g. in $pp$ or $e^+e^-$ scattering.

The standard relation between Wigner functions and density matrices yields

\begin{equation}\label{charfu}
  \rho(\textbf{K},\textbf{q}) = \rho(\textbf{K},\textbf{0})\int\!\!d^3X\;p(\textbf{X}|\textbf{K})e^{-i\textbf{q}\cdot
  \textbf{X}},
\end{equation}
where

\begin{equation}\label{defiKq}
  \textbf{K} = \frac{1}{2}(\textbf{p}_1 + \textbf{p}_2),\qquad \textbf{q} = \textbf{p}_1 -\textbf{p}_2
\end{equation}
and $\rho(\textbf{K},\textbf{q})$ is the density matrix corresponding to the
Wigner function $W(\textbf{X},\textbf{K})$.

Using the HBT method one gets information about the density matrix
$\rho(\textbf{K},\textbf{q})$ from the measured two-, and sometimes more-,
particle correlation functions. In most models, when final state interactions
including resonance production are neglected or corrected for, the basic
formula is \cite{KAR}

\begin{equation}\label{karczm}
  P(\textbf{p}_1,\ldots,\textbf{p}_n) = C_n\sum_P\prod_{j=1}^n\rho(\textbf{p}_j,\textbf{p}_{Pj}),
\end{equation}
where $P(\textbf{p}_1,\ldots,\textbf{p}_n)$ is the $n$-particle probability
distribution, the summation is over the $n!$ permutations of the indices $j$
and $C_n$ are normalization constants.

When this density matrix is known, the profiles of the homogeneity regions
$p(\textbf{X}|\textbf{K})$ can be obtained by inverting the Fourier
transformation. There are, however, ambiguities in the determination of the
density matrix from the data. The fits to the data do not change under the
transformations \cite{BIZ}, \cite{ZAL3}

\begin{equation}\label{ambrho}
  \rho(\textbf{K},\textbf{q}) \rightarrow
  e^{if(\textbf{p}_1)}\rho(\textbf{K},\textbf{q})e^{-if(\textbf{p}_2)},
\end{equation}
where $f$ can be any real-valued function. They are also invariant under
complex conjugation of the density matrix, but this is less interesting since
it corresponds just to a space inversion of the homogeneity region. Conversely,
(\ref{ambrho}) and complex conjugation are the only transformations of the
density matrix which leave the probabilities (\ref{karczm}) unchanged
\cite{ZAL3}. The purpose of the present paper is to quantify the ambiguities in
the profiles $p(\textbf{X}|\textbf{K})$ due to the ambiguity (\ref{ambrho}).

\section{Results}

In the terminology of the calculus of probabilities, relation (\ref{charfu})
means that, at given $\textbf{K}$,
$\rho(\textbf{K},\textbf{q})/\rho(\textbf{K},\textbf{0})$  is  the
characteristic function of the distribution $p(\textbf{x}|\textbf{K})$. The
expansion of the logarithm of the characteristic function in powers of the
components of $\textbf{q}$ yields the cumulants ${\cal K}(r_x,r_y,r_z)$ of the
distribution $p(\textbf{x}|\textbf{K})$:

\begin{equation}\label{}
  \log\rho(\textbf{K},\textbf{q}) =
  \sum_{r_x,r_y,r_z}\frac{q_x^{r_x}}{r_x!}\frac{q_y^{r_y}}{r_y!}\frac{q_z^{r_z}}{r_z!}i^r{\cal
  K}(r_x,r_y,r_z) + \log\rho(\textbf{K},\textbf{0}).
\end{equation}
We introduce the notation $r = r_x+r_y+r_z$ and call a cumulant even (odd) when
$r$ is even (odd). Let us rewrite the logarithm of the density matrix in the
form

\begin{equation}\label{}
  \log \rho(\textbf{K},\textbf{q}) = \log\rho_0(\textbf{K},\textbf{q}) + i\chi(\textbf{K},\textbf{q}),
\end{equation}
where both $\chi$ and $\rho_0$ are real. The hermiticity of the density matrix
implies that $\rho_0$ is an even function of $\textbf{q}$ and $\chi$ is an odd
function of $\textbf{q}$. Thus, $\rho_0$ yields the even cumulants and $\chi$
the odd cumulants. Since the ambiguity (\ref{ambrho}) affects only $\chi$, the
even cumulants can be unambiguously found using the HBT method. This is our
first result. A notable case are the $r=2$ cumulants which yield the famous HBT
radii (see e.g. \cite{WIH}). The measurability of the HBT radii has been
already pointed out in \cite{BIZ}. Another useful implication is the
measurability the distribution of distances between pairs of points where
particles with given momentum $\textbf{K}$ are produced. The corresponding
characteristic function is

\begin{equation}\label{chadis}
  \phi(\textbf{q}|\textbf{K}) = \int\!\!d^3X_1d^3X_2 p(\textbf{X}_1|\textbf{K})p(\textbf{X}_2|\textbf{K})
  e^{iq(\textbf{X}_1 - \textbf{X}_2)} =
  \frac{\rho(\textbf{K},\textbf{q})\rho(\textbf{K},-\textbf{q})}{(\rho(\textbf{K},\textbf{0}))^2}.
\end{equation}
The $\log\phi(\textbf{q}|\textbf{K})$ is an even function of $\textbf{q}$,
therefore, all the odd cumulants vanish, while the even ones are just twice the
measurable cumulants of $p(\textbf{x}|\textbf{K})$. This well-known result is
basic for the method of imaging (cf. e.g. \cite{DAP}). In the present
derivation the smoothness assumption is replaced by the closely related
assumption that formulae (\ref{emiden}) and (\ref{emifun}) can be
simultaneously used.

The general formula for the unmeasurable contributions from function $f$ to the
cumulants is

\begin{equation}\label{}
  \delta {\cal K}(r_x,r_y,r_z) =
  \left(\frac{-i}{2}\right)^{r-1}
  \frac{\partial^rf(\textbf{K})}{\partial K_x^{r_x}\partial K_y^{r_y}\partial K_z^{r_z}}
\frac{1 - (-1)^r}{2}.
\end{equation}

The $r=1$ cumulants give $\langle \textbf{x} \rangle(\textbf{K})$, i.e. the
position of the center of the homogeneity region as a function of $\textbf{K}$.
The unmeasurable parts of these cumulants yields for the uncertainty of
position

\begin{equation}\label{cumone}
  \delta \langle \textbf{x} \rangle(\textbf{K}) = \mbox{{\boldmath$\nabla$}}f(\textbf{K}).
\end{equation}
Formally, this is analogous to the uncertainty of the electromagnetic vector
potential due to the freedom of time-independent gauge transformations. E.g. it
is possible to choose $f(\textbf{K})$ so that for every homogeneity region
$\langle z \rangle = 0$. The difference is, however, that in electrodynamics
the choice of gauge does not affect the physics and is purely a matter of
convenience, while here only one choice of $f(\textbf{K})$, except for an
irrelevant additive constant, is correct and yields the true profiles of the
homogeneity regions. Since experimental momentum distributions give no
information about $f(\textbf{K})$, this function must be deduced from theory or
from a model. As seen from (\ref{cumone}), it is enough to fix the function
$\langle \textbf{x} \rangle(\textbf{K})$, because then $\delta\langle
\textbf{x} \rangle(\textbf{K}) \equiv \textbf{0}$, $f$ reduces to a constant
and all the cumulants can be unambiguously calculated. Consequently, the full
profiles $p(\textbf{x}|\textbf{K})$ are measurable. It should be kept in mind
that changing $f(\textbf{K})$ does not mean only shifting the homogeneity
regions with respect to each other. In general, the higher cumulants also
change and consequently the homogeneity regions are not only shifted, but also
deformed. A one-dimensional example where the change affects the first and
third order cumulants has been discussed and illustrated by a figure in
\cite{BIZ}.

From the two-body correlation function one can obtain
$\rho_0(\textbf{K},\textbf{q})$ - the absolute value of the density matrix.
Often it is assumed, usually implicitly, that $\rho_0$ can be used as an
approximation for the true density matrix $\rho$. As seen from our analysis,
$\rho_0$ yields correctly all the even cumulants and thus embodies all the
information contained in the distribution of distances between the points where
the hadrons have been produced. Replacing $\rho$ by $\rho_0$ implies also,
however, putting arbitrarily all the odd cumulants equal zero. This may lead to
disagreement with experiment for the three-particle correlation function (cf.
e.g. the review \cite{WIH} ). If the three-particle correlation function
happens to be reproduced correctly and if assumption (\ref{karczm}) holds,
there will be also agreement with experiment for all the more-particle
correlation functions \cite{ZAL3}. The profiles may be still wrong, however,
because of the ambiguity (\ref{ambrho}).

\section{Conclusions}

Let us summarize our conclusions. All the even cumulants of the profiles
$p(\textbf{x}|\textbf{K})$ are unambiguously measurable.  This implies in
particular the measurability of the HBT radii and of the distributions of
distances between points where the identical particles with the same momenta
are produced. These results are known, but we give a new, very simple proof
which clearly exhibits the necessary assumptions.

The odd cumulants, though separately not measurable, are strongly correlated
with each other. In particular, it is enough to know the positions of the
centers of all the homogeneity regions $\langle \textbf{x} \rangle
(\textbf{K})$, in order to reconstruct the full profiles
$p(\textbf{x}|\textbf{K})$. Our analysis justifies the following recipe.

Prepare as input: the data on the single-particle momentum distributions, the
two-body and the three-body correlation functions and a distribution $\langle
\textbf{x} \rangle(\textbf{K})$ assumed to be correct. More-particle
correlation functions could be used to reduce the statistical errors, but in
principle they are redundant. Use any model which reproduces correctly the
experimental input to find a density matrix $\rho_{in}(\textbf{K},\textbf{q})$.
This will be one of the infinitely many density matrices giving exactly the
same fit to the data, but in general very different profiles of the homogeneity
regions. Use this matrix to calculate the three components of the first moment
$\langle \textbf{x} \rangle_{in}(\textbf{K})$. The subscript $in$ is used to
distinguish the moments given by the model from the true moments contained in
the input. Calculate the correction

\begin{equation}\label{}
  \delta \langle \textbf{x} \rangle
(\textbf{K}) = \langle \textbf{x} \rangle(\textbf{K}) - \langle \textbf{x}
\rangle_{in} (\textbf{K})
\end{equation}
Substitute it into (\ref{cumone}) and solve for $f(\textbf{K})$. Correct the
density matrix $\rho_{in}$ according to (\ref{ambrho}). Substitute the
corrected density matrix into (\ref{charfu}) and solve for
$p(\textbf{x}|\textbf{K})$ by inverting the Fourier transformation.

\vspace{0.5cm} \Large{\textbf{Acknowledgements}}\normalsize

The author thanks A. Bialas for helpful comments.

\end{document}